\documentstyle[12pt,worldsci]{article}
\begin{document}
\title{A PROOF OF QUARK CONFINEMENT IN QCD
\footnote{Report-no: CHIBA-EP-107 (hep-th/9808186); Talk given at the
3rd International conference on Quark Confinement and the Hadron
Spectrum, 7-12 June 1998, Jefferson Lab., Newport News, VA, USA.}
}
\author{Kei-Ichi Kondo\\
{\em Department of Physics, Faculty of Science, Chiba University,
Chiba 263, JAPAN}\\
}
\maketitle
\setlength{\baselineskip}{2.6ex}

\vspace{0.7cm}
\begin{abstract}

I propose to reformulate the gauge field theory as the
perturbative deformation of a novel topological quantum field
theory.  It is shown that this reformulation leads to quark
confinement in QCD$_4$.  Similarly, the fractional charge
confinement is also derived in the strong coupling phase of
QED$_4$.  As a confinement criterion, we use the area decay of the
expectation value of the Wilson loop.

\end{abstract}
\vspace{0.7cm}

\section{Introduction}

  An idea of magnetic monopole and its condensation was used in the
analytical proofs of "quark" confinement so far: 

\begin{enumerate}
\item[]
1975: Polyakov\cite{Polyakov75} for 3D compact U(1) gauge theory 
\item[]
1977: Polyakov\cite{Polyakov77} for 3D Georgi-Glashow model 
\item[]
1994: Seiberg-Witten\cite{SW94} for 4D N=2 SUSY YM/QCD

\end{enumerate}
 How to prove quark confinement in QCD$_4$ analytically?
Recent numerical simulations have established that also in QCD the
magnetic monopole plays the dominant role in quark confinement
(monopole dominance) under the Abelian projection \cite{tHooft81}.
To explain this fact analytically and then to prove quark
confinement, we propose to use a novel type of topological quantum
field theory (TQFT) \cite{Witten88} which describes 4D magnetic
monopole, and to reformulate QCD as a perturbative
deformation of the TQFT.

We can show:\cite{KondoI,KondoII,KondoIII,KondoIV}
\begin{enumerate}
\item  The TQFT is derived from QCD (without any approximation) in
the maximal Abelian gauge (MAG).  

\item   The calculation of the Wilson loop in 4D YM theory is
reduced to that in 2D nonlinear sigma model (NLSM) due to the
dimensional reduction.  This is an exact result.

\item  Area decay of the Wilson loop is derived from 2D Instanton
calculus in NLSM.
\end{enumerate}

\section{Basic Idea}

We take into account the topological nontrivial background
field $\Omega_\mu$ which corresponds to 4D magnetic monopole current
$k_\mu$.  The current $k_\mu$ obeys the topological  conservation
law $\partial_\mu k_\mu = 0$ and hence
$k_\mu$ denotes a loop in 4D spacetime.
 Usually, $\Omega_\mu$ denotes an arbitrary, but fixed background
(as a solution of classical field equation).  In our case, we sum up
all topological nontrivial background
$\Omega_\mu$ after integrating out quantum fluctuations $Q_\mu$,
\begin{eqnarray}
  Z_{QCD} = \sum_{\Omega} \int [dQ_\mu]
 \exp\{-S_{QCD}[\Omega_\mu+Q_\mu] \} , 
\end{eqnarray}
where
\begin{eqnarray}
 S_{QCD}   &=& \int d^Dx  \left[ -{1 \over 2g^2} 
 {\rm tr}_G({\cal F}_{\mu\nu}{\cal F}_{\mu\nu})
 + \bar \psi (i \gamma^\mu {\cal D}_\mu[{\cal A}] - m) \psi \right] .
\end{eqnarray}
We must regard this background to be different from the 4D instanton
discussed in the topological YM theory (Witten\cite{Witten88}) and
also from 3D magnetic monopole (Polyakov\cite{Polyakov77}).

\section{Quantization of YM theory based on BRST formalism}

Using the nilpotent BRST transformation $\delta_B$, it is well
known that the gauge fixing and the Faddeev-Popov (FP) ghost part is
written in the form,
\begin{eqnarray}
  {\cal L}_{gf} = - i \delta_B G_{gf}, 
\quad
 G_{gf} := {\rm tr}_{G} \left\{
\bar C \left( F[{\cal A}] + {\alpha \over 2} \phi \right) \right\} , 
\end{eqnarray}
where $\alpha$ is the gauge-fixing parameter and $\phi$ is the
Lagrange-multiplier field. 
For the manifestly covariant Lorentz gauge, we choose 
$F[{\cal A}]=\partial^\mu {\cal A}_\mu$.
\par
In the maximal Abelian gauge (MAG),\cite{KondoI,KondoII} a suitable
choice of gauge fixing parameter $\alpha=-2$ leads to 
\begin{eqnarray}
  G_{gf} :=  \bar \delta_B {\rm tr}_{G/H}
\left({1 \over 2}{\cal A}_\mu(x) {\cal A}_\mu(x) + i C(x) \bar C(x)
\right) .
\end{eqnarray}
where $\bar \delta_B$ is the (nilpotent) anti-BRST transformation.
Using the decomposition,
\begin{eqnarray}
  {\cal A}_\mu(x) = \Omega_\mu(x) + Q_\mu(x)
= {i \over g} U(x) \partial_\mu U(x)^\dagger 
+ U(x) {\cal V}_\mu(x) U^\dagger(x) ,
\end{eqnarray}
the action of the TQFT is given by
\begin{eqnarray}
  S_{TQFT} := \int d^Dx \ i \delta_B \bar \delta_B \ {\rm tr}_{G/H}
\left({1 \over 2}\Omega_\mu(x) \Omega_\mu(x) + i C(x) \bar C(x)
\right), 
\label{TFT}
\end{eqnarray}
and the partition function is given by
\begin{eqnarray}
  Z_{TQFT} = \int [dU][dC][d\bar C][d\phi]
 \exp\{-S_{TQFT}[\Omega_\mu] \} , 
\end{eqnarray}
where $U$ is an element of the gauge group $G$.

\section{Strategy of a proof}

We consider D-dim. QCD (with a gauge group G) for $D >2$.

\subsection{Step 1:\cite{KondoII}}
In MAG, D-dim. QCD (QCD$_D$) is reformulated as a perturbative
deformation of TQFT$_D$.  MAG is a partial gauge fixing which fixes
the coset part G/H and leaves the maximal torus group H invariant.

\subsection{Step 2:\cite{KondoII}}
It is shown that TQFT$_D$ is equivalent to the coset G/H nonlinear
sigma model (NLSM) in (D-2) dimensions.  This is a consequence of
Parisi-Sourlas dimensional reduction due to the hidden supersymmetry
in TQFT (\ref{TFT}).
As extensively discussed more than 20 years ago, 
QCD$_4$ and NLSM$_2$ have common properties: 
renormalizability, asymptotic freedom ($\beta(g) < 0$),
dynamical mass generation, existence of instanton solution, no phase
transition for any value of coupling constant (one phase),
etc. 
This similarity between two theories can be understood from this
correspondence,  
\begin{eqnarray}
 QCD_4 \supset TQFT_4 \Longleftrightarrow G/H~ NLSM_2
\end{eqnarray}
For  $G=SU(2)$, G/H NLSM is equivalent to O(3) NLSM.
Existence of 2D instanton is guaranteed for any $N$, because
$\Pi_2(SU(N)/U(1)^{N-1})=\Pi_1(U(1)^{N-1})={\bf Z}^{N-1}$.

\subsection{Step 3:\cite{KondoIV}}
We define the full string tension $\sigma$ by
\begin{eqnarray}
  \sigma := - \lim_{A(C) \rightarrow \infty}
  {1 \over A(C)} \ln \langle W^C[{\cal A}] \rangle_{YM_4} ,
\end{eqnarray}
where $W^C[{\cal A}]$ is the non-Abelian Wilson loop with the area 
$A(C)$, 
\begin{eqnarray}
 W^C [{\cal A}] :=  {\rm tr} \left[ {\cal P} 
 \exp \left( i q \oint_C {\cal A}_\mu^A(x) T^A dx^\mu
 \right) \right] .
\end{eqnarray}
On the other hand, the diagonal string tension $\sigma_{Abel}$ is
defined by
\begin{eqnarray}
  \sigma_{Abel} 
   := -  \lim_{A(C) \rightarrow \infty} {1 \over A(C)} \ln  
  \left\langle  W^C[a^\Omega]
 \right\rangle_{TFT_4} ,
\end{eqnarray}
using the diagonal Wilson loop,
\begin{eqnarray}
  W^C[a^\Omega] =  \exp \left( i q \oint_C 
  dx^\mu a_\mu^\Omega (x) \right),
  \quad a_\mu^\Omega(x) := \Omega_\mu^3(x) 
:= {\rm tr}(T^3 \Omega_\mu(x)) .
  \label{dWl}
\end{eqnarray}
From the dimensional reduction, we find
\begin{eqnarray}
 \langle W^C[a^\Omega] \rangle_{TFT_4}
 =  \left\langle  W^C[a^\Omega]
 \right\rangle_{NLSM_2} .
\end{eqnarray}
Then it is shown that in the large Wilson loop limit
\begin{eqnarray}
  |\sigma - \sigma_{Abel}| \downarrow 0 \quad
(A(C) \uparrow \infty) .
\end{eqnarray}
This is derived from the non-Abelian Stokes theorem.
For the large (non-intersecting planar) Wilson loop, the full string
tension $\sigma$ is saturated by the diagonal string tension
$\sigma_{Abel}$.  This explains the Abelian dominance and magnetic
monopole dominance.

\subsection{Step 4:\cite{KondoII}}
The whole problem is reduced to calculating 
\begin{eqnarray}
 \left\langle  W^C[a^\Omega]
 \right\rangle_{NLSM_2} 
 = \left\langle    e^{i {q \over g} 8\pi Q_S } 
 \right\rangle_{NLSM_2}  ,
\quad
  Q_S = {1 \over  8\pi } \int_S d^2z \ \epsilon_{\mu\nu}
{\bf n} \cdot (\partial_\mu {\bf n} \times \partial_\nu {\bf n})  .
\end{eqnarray}
Note that the integrand of $Q_S$ is the instanton density in
NLSM$_2$.  Therefore, $Q_S$ counts the number of instantons minus
that of anti-instantons inside the area $S$ bounded by the Wilson
loop
$C$. In order to perform the actual calculation, we adopt the the
naive instanton calculus (dilute gas approximation).  This leads to 
the area law of the diagonal Wilson loop and the
non-zero diagonal string tension
$\sigma_{Abel}$ for the fractional charge $q$ (i.e., when $q/g$ is
not an integer).  More systematic instanton calculations enable us
to identify the instanton solution with the Coulomb
gas of vortices.

Note that all the above steps are exact except for the instanton
calculus of the Wilson loop in NLSM.
The above results imply that the non-zero string tension $\sigma$ in
QCD$_4$ follows from the non-zero diagonal string tension $\sigma_A$
in NLSM$_2$.  The problem of proving area law in QCD$_4$ is
reduced to the equivalent problem in NLSM$_2$.

\par
\vskip 0.5cm
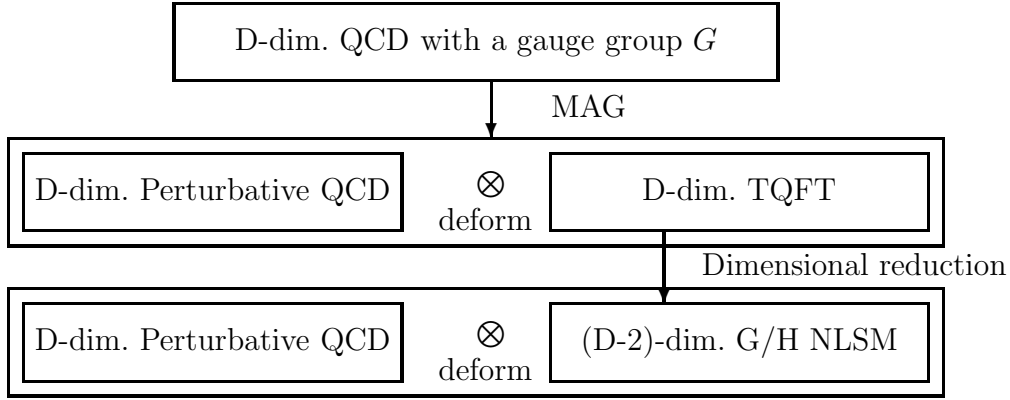
\begin{center}
\unitlength=1cm
\thicklines
 \begin{picture}(12,6)
 \put(2,5){\framebox(8,1){D-dim. QCD with a gauge group $G$}}
 \put(6.2,5){\vector(0,-1){0.8}}
 \put(7,4.5){MAG}
 \put(-0.2,2.8){\framebox(12.4,1.4){}}
 \put(0,3){\framebox(5,1){D-dim. Perturbative QCD}}
 \put(6,3.5){$\bigotimes$}
 \put(5.5,3){deform}
 \put(7,3){\framebox(5,1){D-dim. TQFT}}
 \put(8.5,3){\vector(0,-1){1}}
 \put(9,2.4){Dimensional reduction}
 \put(-0.2,0.8){\framebox(12.4,1.4){}}
 \put(0,1){\framebox(5,1){D-dim. Perturbative QCD}}
 \put(6,1.5){$\bigotimes$}
 \put(5.5,1){deform}
 \put(7,1){\framebox(5,1){(D-2)-dim. G/H NLSM}}
 \end{picture}
\end{center}

Similar strategy is also applied to QED$_4$ to prove the existence of
strong coupling confinement phase.\cite{KondoIII}  This follows from
the existence of Berezinski-Kosterlitz-Thouless transition of the O(2)
NLSM$_2$.
%

\end{document}